\documentclass[12pt,hyper]{JHEP3}
\usepackage{graphics,psfrag}
\usepackage{amsmath,amsthm,amssymb,epsfig,euscript,array,cite}
\usepackage{cases,empheq,pmat}
\newtheorem{thm}{Theorem}[section]
\newtheorem{lem}[thm]{Lemma}
\newtheorem{prop}[thm]{Proposition}
\newtheorem*{cor}{Corollary}

\theoremstyle{definition}
\newtheorem{defn}{Definition}[section]

\theoremstyle{remark}

\newcounter{multieqs}

\newcommand{\be}{\begin{equation}}
\newcommand{\ee}{\end{equation}}
\newcommand{\eq}[1]{(\ref{#1})}
\newcommand{\bit}{\begin{itemize}}  \newcommand{\eit}{\end{itemize}}

\newcommand{\bm}[1]{\mbox{\boldmath $#1$}}
\newcommand{\rf}[1]{(\ref{#1})}

\def\bd{\begin{document}}
\def\ed{\end{document}}
\def\nn{\nonumber}
\def\bea{\begin{eqnarray}}
\def\eea{\end{eqnarray}}
\let\bm=\bibitem

\def\la{\langle}
\def\ra{\rangle}

\def\npb#1#2#3{Nucl. Phys. {\bf{B#1}} #3 (#2)}
\def\plb#1#2#3{Phys. Lett. {\bf{#1B}} #3 (#2)}
\def\prl#1#2#3{Phys. Rev. Lett. {\bf{#1}} #3 (#2)}
\def\prd#1#2#3{Phys. Rev. {D \bf{#1}} #3 (#2)}
\def\cmp#1#2#3{Comm. Math. Phys. {\bf{#1}} #3 (#2)}
\def\cqg#1#2#3{Class. Quantum Grav. {\bf{#1}} #3 (#2)}
\def\nppsa#1#2#3{Nucl. Phys. B (Proc. Suppl.) {\bf{#1A}}#3 (#2)}
\def\ap#1#2#3{Ann. of Phys. {\bf{#1}} #3 (#2)}
\def\ijmp#1#2#3{Int. J. Mod. Phys. {\bf{A#1}} #3 (#2)}
\def\rmp#1#2#3{Rev. Mod. Phys. {\bf{#1}} #3 (#2)}
\def\mpla#1#2#3{Mod. Phys. Lett. {\bf A#1} #3 (#2)}
\def\jhep#1#2#3{J. High Energy Phys. {\bf #1} #3 (#2)}
\def\atmp#1#2#3{Adv. Theor. Math. Phys. {\bf #1} #3 (#2)}

\def\N{{\cal N}}
\def\sst{\scriptscriptstyle}
\def\thetabar{\bar\theta}
\def\Tr{{\rm Tr}}
\def\one{\mbox{1 \kern-.59em {\rm l}}}

\def\a{\alpha}      \def\da{{\dot\alpha}}  \def\dA{{\dot A}}
\def\b{\beta}       \def\db{{\dot\beta}}  
\def\g{\gamma}  \def\G{\Gamma}  \def\dc{{\dot\gamma}}  
\def\d{\delta}  \def\D{\Delta}  \def\ddt{\dot\delta}  
\def\e{\epsilon}        \def\ve{\varepsilon}  
\def\f{\phi}    \def\F{\Phi}    \def\vvf{\f}  
\def\h{\eta}  
\def\k{\kappa}  
\def\l{\lambda} \def\L{\Lambda}  
\def\m{\mu} \def\n{\nu}  
\def\o{\omega}  
\def\p{\pi} \def\P{\Pi}  
\def\r{\rho}  
\def\s{\sigma}  \def\S{\Sigma}  
\def\t{\tau}  
\def\th{\theta} \def\Th{\Theta} \def\vth{\vartheta}  
\def\X{\Xeta}  
\def\z{\zeta}  

\def\na{\nabla}  

\def\cA{{\cal A}} \def\cB{{\cal B}} \def\cC{{\cal C}}  
\def\cD{{\cal D}} \def\cE{{\cal E}} \def\cF{{\cal F}}  
\def\cG{{\cal G}} \def\cH{{\cal H}} \def\cI{{\cal I}}  
\def\cJ{{\cal J}} \def\cK{{\cal K}} \def\cL{{\cal L}}  
\def\cM{{\cal M}} \def\cN{{\cal N}} \def\cO{{\cal O}}  
\def\cP{{\cal P}} \def\cQ{{\cal Q}} \def\cR{{\cal R}}  
\def\cS{{\cal S}} \def\cT{{\cal T}} \def\cU{{\cal U}}  
\def\cV{{\cal V}} \def\cW{{\cal W}} \def\cX{{\cal X}}  
\def\cY{{\cal Y}} \def\cZ{{\cal Z}}

\def\ua{\underline{\alpha}}  
\def\uc{\underline{\phantom{\alpha}}\!\!\!\gamma}  
\def\um{\underline{\mu}}  
\def\ud{\underline\delta}  
\def\ue{\underline\epsilon}  
\def\una{\underline a}\def\unA{\underline A}  
\def\unb{\underline b}\def\unB{\underline B}  
\def\unc{\underline c}\def\unC{\underline C}  
\def\und{\underline d}\def\unD{\underline D}  
\def\une{\underline e}\def\unE{\underline E}  
\def\unf{\underline{\phantom{e}}\!\!\!\! f}\def\unF{\underline F}  
\def\unm{\underline m}\def\unM{\underline M}  
\def\unn{\underline n}\def\unN{\underline N}  
\def\unp{\underline{\phantom{a}}\!\!\! p}\def\unP{\underline P}  
\def\unq{\underline{\phantom{a}}\!\!\! q}  
\def\unQ{\underline{\phantom{A}}\!\!\!\! Q}  
\def\unH{\underline{H}}  
  
\def\As {{A \hspace{-6.4pt} \slash}\;}  
\def\bs {{b \hspace{-6.4pt} \slash}\;}  
\def\Ds {{D \hspace{-6.4pt} \slash}\;}  
\def\ds {{\del \hspace{-6.4pt} \slash}\;}  
\def\ss {{\s \hspace{-6.4pt} \slash}\;}  
\def\ks {{ k \hspace{-6.4pt} \slash}\;}  
\def\ps {{p \hspace{-6.4pt} \slash}\;}   
\def\xs {{x \hspace{-6.4pt} \slash}\;}  
\def\pas {{{p_1} \hspace{-6.4pt} \slash}\;}  
\def\pbs {{{p_2} \hspace{-6.4pt} \slash}\;}   
\def\cFs {{{\cal F} \hspace{-6.4pt} \slash}\;}

\def\Dh{{\hat{D}}}
\def\Gh{{\hat{G}}}
\def\Fh{{\hat{F}}}
\def\Ih{{\hat{I}}} 
\def\Jh{{\hat{J}}} 
\def\Kh{{\hat{K}}}
\def\Lh{{\hat{L}}} 
\def\Ph{{\hat{P}}}
\def\Rh{{\hat{R}}}
\def\Vh{{\hat{V}}} 
\def\Xh{{\hat{X}}}
 
\def\ah{{\hat{\a}}}
\def\bh{{\hat{\b}}}
\def\gh{{\hat{\g}}}
\def\dh{{\hat{\d}}}
\def\hh{\hat{h}}
\def\uh{\hat{u}}  
\def\xh{\hat{x}}  
\def\yh{\hat{y}}  
\def\ph{\hat{p}}  
\def\xih{\hat{\xi}}  
\def\chih{\hat{\chi}}

\def\psit{\tilde{\psi}}  
\def\Psit{\tilde{\Psi}}   
\def\Psibt{\tilde{\bar{Psi}}}  

\def\st{\tilde{\sigma}}  

\def\Phit{\tilde{\Phi}}   
\def\Phitb{\overline{\tilde{Phi}}}  
\def\tht{\tilde{\th}}  
\def\lt{\tilde{\l}}
\def\chit{\tilde{\chi}}   
\def\phit{\tilde{\phi}} 

\def\At{\tilde{A}}
\def\Bt{\tilde{B}}
\def\Ct{\tilde{C}}
\def\Dt{\tilde{D}}
\def\Et{\tilde{E}}
\def\Ft{\tilde{F}}
\def\Ht{\tilde{H}}
\def\It{\tilde{I}}
\def\Jt{\tilde{J}}
\def\Qt{\tilde{Q}}  
\def\Rt{\tilde{R}}  
\def\Mt{\tilde{M }}  
\def\Nt{\tilde{N}}   
\def\St{\tilde{S}}
\def\Vt{\tilde{V}}
\def\Xt{\tilde{X}} 
\def\at{\tilde{a}}
\def\ct{\tilde{c}}   
\def\htt{\tilde{h}} 
\def\ft{\tilde{f}}
\def\gt{\tilde{g}}
\def\pt{\tilde{p}}  
\def\qt{\tilde{q}}  
\def\vt{\tilde{v}}  
\def\nt{\tilde{n}}  
\def\ut{\tilde{u}}  
\def\wt{\tilde{w}}  
\def\zt{\tilde{z}} 
\def\xt{\tilde{x}} 
\def\yt{\tilde{y}} 
\def\Psit{\tilde{\Psi}}
\def\vphit{\tilde{\varphi}}  

\def\delb{\bar{\partial}}  
\def\thb{\bar{\theta}}
\def\mub{\bar{\mu}}
\def\lamb{\bar{\l}}
\def\psib{\bar{\psi}}
\def\sb{\bar{\sigma}}
\def\xib{\bar{\xi}}
\def\chib{\bar{\chi}}

\def\Phib{\bar{\Phi}}
\def\Lamb{\bar{\Lambda}}
\def\Sb{{\overline \Sigma}}
\def\cb{\bar{c}}
\def\hb{\bar{h}}
\def\qb{\bar{q}}
\def\wb{\bar{w}}
\def\ub{\bar{u}}
\def\zb{{\bar{z}}}
\def\Hb{\bar{H}}
\def\Qb{{\bar Q}}

\def\Ab{{\overline A}} \def\Bb{{\overline B}} \def\Cb{{\overline C}}  
\def\Db{{\overline D}} \def\Eb{{\overline E}} \def\Fb{{\overline F}}  
\def\Gb{{\overline G}} 
\def\Ib{{\overline I}}  
\def\Jb{{\overline J}} \def\Kb{{\overline K}} \def\Lb{{\overline L}}  
\def\Mb{{\overline M}} \def\Nb{{\overline N}} \def\Ob{{\overline O}}  
\def\Pb{{\overline P}}  \def\Rb{{\overline R}}  
 \def\Tb{{\overline T}} \def\Ub{{\overline U}}  
\def\Vb{{\overline V}} \def\Wb{{\overline W}} \def\Xb{{\overline X}}  
\def\Yb{{\overline Y}} \def\Zb{{\overline Z}}  

\def\fb{{\overline f}}
\def\gb{{\overline g}}
\def\mb{{\overline m}}
\def\lb{{\overline l}}
\def\yb{{\overline y}}

\def\ba{{\bf a}} 
\def\bk{{\bf k}}  
\def\bl{{\bf l}}  
\def\bp{{\bf p}}  
\def\bq{{\bf q}}  
\def\br{{\bf r}}
\def\bt{{\bf t}}
\def\bu{{\bf u}}
\def\bv{{\bf v}}
\def\bx{{\bf x}}  
\def\by{{\bf y}}  
\def\bR{{\bf R}}  
\def\bV{{\bf V}}

\def\bone{{\bf 1}}  

\def\va{{\vec a}}
\def\vk{{\vec k}}
\def\vp{{\vec p}}
\def\vq{{\vec q}}
\def\vx{{\vec x}}
\def\vy{{\vec y}}
\def\vu{{\vec u}}
\def\vv{{\vec v}}

\def\vs{{\vec \sigma}}
\def\vtau{{\vec \tau}}

\newcommand{\ov}[1]{\overrightarrow{#1}}

\def\frA{\mathfrak{A}}
\def\frB{\mathfrak{B}}
\def\frC{\mathfrak{C}}
\def\frD{\mathfrak{D}}
\def\frE{\mathfrak{E}}
\def\frF{\mathfrak{F}}
\def\frG{\mathfrak{G}}
\def\frH{\mathfrak{H}}
\def\frM{\mathfrak{M}}
\def\frN{\mathfrak{N}}
\def\frR{\mathfrak{R}}
\def\frW{\mathfrak{W}}

\def\fra{\mathfrak{a}}
\def\frb{\mathfrak{b}}
\def\frf{\mathfrak{f}}
\def\frg{\mathfrak{g}}
\def\frh{\mathfrak{h}}
\def\frl{\mathfrak{l}}
\def\frs{\mathfrak{s}}
\def\fri{\mathfrak{i}}
\def\frj{\mathfrak{j}}

\def\ma{\mathfrak{a}}
\def\mg{\mathfrak{g}}
\def\mh{\mathfrak{h}}
\def\mR{\mathfrak{R}}
\def\mN{\mathfrak{N}}

\def\d{\delta}\def\D{\Delta}\def\ddt{\dot\delta}  
  
\def\pa{\partial} \def\del{\partial}  
\def\xx{\times}  
\def\uno{\mbox{1 \kern-.59em {\rm l}}}    
  
\def\trp{^{\top}}  
\def\inv{^{-1}}  
\def\dag{{^{\dagger}}}  
\def\pr{^{\prime}}  
  
\def\rar{\rightarrow}  
\def\lar{\leftarrow}  
\def\lrar{\leftrightarrow}  
  
\newcommand{\0}{\,\!}      
\def\one{1\!\!1\,\,}  
\def\im{\imath}  
\def\jm{\jmath}  
  
\newcommand{\tr}{\mbox{tr}}  
\newcommand{\slsh}[1]{/ \!\!\!\! #1}  
  
\def\vac{|0\rangle}  
\def\lvac{\langle 0|}  
  
\def\hlf{\frac{1}{2}}  
\def\ove#1{\frac{1}{#1}}  

\def\Box{\square}  
\def\CC {\mathbb{C}}
\def\FF {\mathbb{F}}
\def\RR{\mathbb{R}}
\def\NN{\mathbb{N}}  
\def\ZZ{\mathbb{Z}}  
\def\bb#1{{\bf #1}}  
\def\bcomment#1{}  
\def\bfhat#1{{\bf \hat{#1}}}  
\def\VEV#1{\left\langle #1\right\rangle}  

\newcommand{\ex}[1]{{\rm e}^{#1}} \def\ii{{\rm i}}  

\newcommand{\lrbrk}[1]{\left(#1\right)}
\newcommand{\sfrac}[2]{{\textstyle\frac{#1}{#2}}}
 
\def\stw{{\sqrt{2}}}

\def\rf {{\rm f}}
\def\ri {{\rm i}}
\def\rj {{\rm j}}
\def\rk {{\rm k}}
\def\rl {{\rm l}}
\def\rs {{\scriptscriptstyle \rm S}}
\def\rt {{\scriptscriptstyle \rm T}}

\def\rQ {{\scriptscriptstyle \rm \cQ}}
\def\rR {{\scriptscriptstyle \rm \cR}}

\def\cQb{{\cal \Qb}}
\def\cRb{{\cal \Rb}}
\def\cWb{{\cal \Wb}}

\def\fd {{\rm N}}
\def\afd {{\overline{\rm N}}}

\def \II {I\hspace{-.1em}I\hspace{.1em}}
\def \IIA {\mbox{\II A\hspace{.2em}}}
\def \IIB {\mbox{\II B\hspace{.2em}}}
\def \gs {g^s}
\def \ls {\lambda^s}

\def \I {{\cal I}}
\def \qs {q\hspace{-.53em}/\hspace{.15em}}
\def \ks {k\hspace{-.53em}/\hspace{.15em}}
\def \YM {{\mbox{\tiny YM}}}
\def \gym {g_{\YM}}

\def \Lc {\L_c}
\def\IR{\relax{\rm I\kern-.18em R}}
\def \id {{\bf 1}}

\def\cci{\ell}
\def\ccj{\ell'}

\author{Chong-Sun Chu  \\  
Centre for Particle Theory
and Department of Mathematics, 
Durham University, Durham, DH1 3LE, UK \\
E-mail:  
\email{chong-sun.chu@durham.ac.uk} }

\title {Cartan-Weyl 3-algebras and the BLG Theory II:\\
Strong-Semisimplicity and Generalized Cartan-Weyl 3-algebras}

\abstract{ 
One of the most important questions in the Bagger-Lambert-Gustavsson (BLG) 
theory
of multiple M2-branes is the choice of the Lie 3-algebra. The Lie 3-algebra 
should be
chosen such that the corresponding BLG 
model is unitary and admits fuzzy 3-sphere as a solution. 
In this paper we propose a new condition: 
the Lie 3-algebras of use must be connected to 
the semisimple Lie algebras describing the gauge symmetry of D-branes via a
certain reduction condition. 
We show that this reduction condition 
leads to a natural generalization of the Cartan-Weyl 
3-algebras  introduced in 
\href{http://arxiv.org/abs/1004.1397}{\cite{cat1}}. 
Similar to a Cartan-Weyl 3-algebra, 
a generalized Cartan-Weyl 3-algebra processes 
a set of step generators characterized by non-degenerate 
roots. However, its Cartan subalgebra is non-abelian in general.
We give
reasons why having a non-abelian Cartan subalgebra may be 
just right to allow for fuzzy 3-sphere solution in the  
corresponding BLG models. We propose that
generalized Cartan-Weyl 3-algebras is the right class of metric Lie 
3-algebras to be used in the BLG theory.

}
 
\preprint{DCPT-10/07}
\keywords{D-Branes, M-Theory, 
Gauge symmetry, Lie $n$-algebra}

\begin{document} 

\section{Introduction}

This paper is a continuation of our paper \cite{cat1}.
In \cite{cat1}, 
we have introduced the concept of a  {\it Cartan-Weyl
3-algebra}. A Lie 3-algebra is Cartan-Weyl if it admits a
Cartan-Weyl basis of generators which consists of 
Cartan subalgebra of mutually commuting generators and 
a set of step generators  that are characterized by a root space
of degenerate 2-forms.
We have also analysed the consistency conditions arising from the
fundamental identity and obtained a  complete classification of
Cartan-Weyl 3-algebras. 

In the case of Lie algebras, the existence of a Cartan-Weyl basis is
equivalent to the fact that the Lie algebra is semisimple, i.e. the Lie
algebra has no nonzero solvable ideals.  For Lie 3 (or higher
$n$)-algebras,  due
to the possibilities of having a number of  different notions of
solvability and semisimplicity (see section 3 below), 
it is not clear whether any of these
notions of semisimplicity is a good characterization of a 
Cartan-Weyl 3-algebra. One of the motivation
of this paper is to address this question.

From the physicist point of view, 
the use of semisimple Lie algebras is both natural and
extremely important since semisimple Lie algebras
are the Lie algebras of compact connected Lie groups, 
natural devices for describing continuous symmetries.
The story is much less clear for Lie 3 (or higher $n$)-algebras 
since the corresponding concepts of
an exponential map or a finite transformation have not been developed. As
a result, it is not clear whether any of these
existing notions of semisimplicity is relevant for the description of multiple
M2-branes in the BLG models \cite{BL1,G1,BL2,BL3}. 
The clarification of this is another motivation 
of this work.

A key observation in our analysis
will be that the theory of root space decomposition for
Lie algebras  
can be readily generalized to Lie 3 (or higher $n$)-algebras.
The root space decomposition  provides the most convenient framework
for analysing the conditions that characterize Cartan-Weyl
3-algebras: the Abelianess of the Cartan subalgebra
and the  non-degeneracy of the root space components. It turns out that
none of the  existing notions of semisimplicity  is 
sufficient to characterize  Cartan-Weyl 3-algebras, and a
stronger notion of semisimplicity is needed. 

In this paper, we will analysis this question and propose a new notion of
semisimplicity.
This new notion of
semisimplicity is motivated by a very natural physical considerations. 
And as it turns out, it leads to a natural generalization of  
Cartan-Weyl 3-algebras, i.e. one which has  a root space whose root
components are non-degenerate and a Cartan subalgebra which is 
non-abelian in general. This generalization is useful because it not
only includes the Cartan Weyl Lie 3-algebra as a sub-case, but it also 
allows us to bypass a
no-go theorem obtained in \cite{cat1},  which states that a
Cartan-Weyl 3-algebra, other than $\cA_4$ itself, 
does not contain the simple Lie 3-algebra 
$\cA_4$ as a subalgebra. As a result, the corresponding BLG theories
do not contain  fuzzy $S^3$ in their (semiclassical)
description \cite{BH,CS2}.  
With the generalized Cartan-Weyl 3-algebras, this is however
possible in general. 
 
The organization of the paper is as follows. In section 2, we 
show that for a metric Lie algebra, the
root decomposition is rather complicated in general. 
What one obtains as a consequence of the existence of an invariant
metric is that the roots with nonzero norm are non-degenerate, see
\eq{rrr}. However further simplifications occur with a semisimple Lie
algebra and one arrives at a Cartan-Weyl basis. In section 3, we 
develop the theory of root decomposition for Lie $n$-algebras. We will
also
examine the various notions of semisimplicity existed in the
literature and show that none of them help to simplify the structure
of the root space decomposition. In section 4, we 
propose a new definition of semisimplicity that is motivated by
consideration of a possible connection between the Lie 3-algebra 
symmetry of multiple M2-branes and the Lie
algebra symmetry of multiple D-branes system. We will show that this notion
of semisimplicity leads to a natural generalization of the
Cartan-Weyl 3-algebras. 
This is the main result of this paper.
A generalized 
Cartan-Weyl 3-algebra is almost the same as a Cartan-Weyl 3-algebra, 
but with the essential difference that its Cartan subalgebra is 
non-abelian in general.
The paper is ended with some further 
discussions in section 5.

\section{Structural Theory of Lie Algebras: Root Decomposition}

\subsection{Root decomposition of Lie algebras}

As a result of it's Lie algebraic structure, a Lie algebra always carry 
a decomposition characterized by  a root space. It turns out the same 
construction can be carried over immediately to  
Lie $n$-algebras. So let us first review the
theory of root space decomposition for Lie algebras.

\subsubsection{Representation of nilpotent Lie algebras}

Let $\rho: \mg \to {\rm gl} (V)$ be a representation of the Lie algebra
$\mg$ and let $\l \in \mg^*$ be any linear form. Define 
\be
V^\l (\mg) := \big\{ v \in V \;|\; 
(\rho(x)- \l(x) E)^m v =0,\; \mbox{for some $m>0$  and for all $x \in  \mg$} 
 \big\},
\ee
where $E$ is the identity operator.
If $V^\l(\mg) \neq 0$, then  $V^\l(\mg)$ is said to be a root
subspace of the representation $\r$, and $\l$ is called a weight. 
A particularly interesting result
is obtained when $\mg$ is nilpotent, 
\begin{thm} 
Let $\rho: \mg \to {\rm gl} (V)$ be a representation of a nilpotent 
complex Lie algebra
$\mg$. Then 
\be \label{decomp1}
V = \bigoplus_{i=1}^s V^{\l_i}(\mg),
\ee
where $\l_i \in \mg^*$ are different {\it weights of the representation
$\r$}. 
\end{thm}

\subsubsection{Weights and roots with respect to a nilpotent subalgebra}

Let $\mg$ be a complex Lie algebra, $\mh$ a nonzero nilpotent
subalgebra of it, and $\r$ a complex linear representation. Applying the
above result to the representation $\r|_\mh$, we obtain the
decomposition
\be \label{decomp2}
V = \bigoplus_{i=1}^s V^{\l_i},
\ee
where the different linear forms $\l_i: \mh \to \CC$ are different
weights of the representation $\r|_\mh$,  and 
$V^{\l_i}$ are the corresponding root subspaces. The space of 
{\it weights of the representation $\r$ with respect to $\mh$} will
be denoted by $\Phi_\r(\mh) = \{ \l_1, \cdots, \l_s\}$. 

This result can be applied to the adjoint representation $\r = \mbox{ad}$ of
$\mg$. The fact that $\mh$ is nilpotent implies that
$\mh \subset \mg^0$ and so $0 \in \Phi_{\rm ad} (\mh)$. Nonzero weights are
called {\it roots of the Lie algebra with respect to $\mh$}. 
Denote the system
of all roots by $\D_\mg(\mh) = \Phi_{\rm ad} (\mh) / \{  0 \}$. 
We obtain:
\begin{thm}\label{decomp3-thm}
Given a nilpotent subalgebra $\mh$ of $\mg$, one has the
decomposition of $\mg$ as a vector space
\be \label{decomp3}
\mg = \mg^0 \oplus \Bigg(\bigoplus_{\a \in \D_\mg(\mh)} \mg^{\a} \Bigg).
\ee
This is called the {\it root decomposition of $\mg$ with respect to $\mh$}.
Here, explicitly it is,
\bea
\mg^0 &=&  
\big\{ y \in \mg \;|\; ({\rm ad}\, x)^m y =0, \; \mbox{for some $m>0$
  and for all $x \in  \mh$} \big\}, \label{g0} \\
\mg^\a &=& \big\{ y \in \mg \;|\; ({\rm ad}\, x - \a(x) E)^m y =0, \; 
\mbox{for some $m>0$  and for all $x \in  \mh$} \big\}.\quad \;\;\;\;
\label{g1}
\eea
\end{thm}
The root spaces $\mg^0, \mg^\a$ are naturally endorsed with a
Lie-algebraic structure. We have:
\begin{prop}
\be\label{gagb}
[ \mg^\a, \mg^\b] \begin{cases} 
\subset \mg^{\a+\b}, & \a+\b \in \Phi_{\rm ad} (\mh),  \\
=0, & \mbox{otherwise}. 
\end{cases}
\ee
Thus the subspace $\mg^0$ is a subalgebra of $\mg$. 
\end{prop}

\subsubsection{Cartan subalgebras}
 
The decomposition \eq{decomp3} of Lie algebra is general and applies for
any nilpotent subalgebra $\mh$ of $\mg$. A particularly interesting
kind of nilpotent subalgebras is the Cartan subalgebra. A subalgebra $\mh$ 
is called a {\it Cartan subalgebra} if it is nilpotent and equal to its
normalizer \footnote{
The normalizer of a subspace $V$ is defined by
$N_\mg (V) =\{ x\in \mg | [x, V] \subset V\}$}. One can easily show that
a Cartan subalgebra is a maximal nilpotent subalgebra but the converse
is not true.

Back to the root decomposition \eq{decomp3}, 
one can show that a nilpotent subalgebra $\mh$ is a Cartan
subalgebra iff $\mg^0 =\mh$. Therefore we obtain:
\begin{thm} \label{root-space-decomp}
Let $\mg$ be a complex Lie algebra and $\mh$ be a  Cartan
subalgebra. One has the root space decomposition
\be \label{decomp4}
\mg = \mh \oplus \Bigg(\bigoplus_{\a \in \D_\mg(\mh)} \mg^{\a} \Bigg).
\ee
\end{thm}
We note on passing the following interesting relation:
\begin{prop}
For the root decomposition \eq{decomp3} of a Lie algebra $\mg$ with respect
to its Cartan subalgebra $\mh$, the Killing metric on $\mh$ can be
expressed in terms of the roots as follows:
\be \label{kroot}
\k(h,k) = \sum_\a n_\a \a(h) \a (k),\quad  \mbox{for $h,k \in \mh$}.
\ee
Here $n_\a:= {\rm dim} \mg^\a \geq 1$. 
\end{prop}
\begin{proof}
This follows from the definition of $\mg^\a$ that, 
when restricted to $\mg^\a$, ${\rm
ad}\, h$, ${\rm ad}\, k$ are of the form
\be \label{adh-adk}
{\rm ad} \, h = \left(\begin{array}{ccc}
\a(h) & & * \\
& \ddots & \\
0 & & \a(h)
\end{array}
\right), \quad
{\rm ad} \, k = \left(\begin{array}{ccc}
\a(k) & & * \\
& \ddots & \\
0 & & \a(k)
\end{array}
\right),
\ee
where the matrices are $n_\a$ dimensional.
\end{proof}

The above discussion leaves open the question of the existence of
Cartan subalgebra. There is in fact a simple way to construct a Cartan
subalgebra if the ground field contains sufficiently many
elements, e.g. for the complex field $\CC$. 
We will refer the readers to any classic textbook,
e.g. \cite{jacobson,OV}, 
for  the details. We will be contented to state here:
\begin{thm}
There always exists a Cartan subalgebra for Lie algebra over $\CC$
\footnote{This is true as long as the ground field $\FF$ is algebraically
closed and is of sufficiently large characteristic.}. Moreover all
Cartan subalgebras of a complex Lie algebra $\mg$ are conjugate under 
automorphism. 
\end{thm}

Although the root decomposition \eq{decomp4} of Lie algebra is quite
neat,  the root spaces $\mg^0$ and $\mg^\a$'s are still
rather complicated in general. It will be  nice to be able to obtain further
information about their structure in special circumstances. 
The simplest situation one can
imagine is that $\mg^\a$ is of  dimension one for $\a \neq 0$ and that all
the $H_I$'s commute with each other. 
In this case, if we denote the generator of $\mg^\a$ by $E^\a$
and the generators of $\mh$ by $H_I$, $I =1, \cdots, N$ for some
$N$, then it follows immediately from \eq{g1} and \eq{gagb} that
\bea \label{cw2}
\; [H_I, H_J] &=&0, \nn\\
\;  [ H_I, E^\a ] &=& \a_I E^\a, \nn\\
\; [E^\a, E^\b] &=& \begin{cases}  
0, & \mbox{if $\a+\b \neq 0$ not  a root},\\
c(\a,\b) E^{\a+ \b}, & \mbox{if $\a+\b \neq 0$ is a root},\\
\in \mh, & \mbox{if $\a+\b =0$}.
\end{cases} 
\eea
for some function $c(\a,\b)$. Such a basis of generators
$\{H_I, E^\a\}$ 
is called a {\it Cartan-Weyl basis} and the Lie algebra is called a 
{\it Cartan-Weyl Lie algebra}. 
The question to ask is  when does such a basis of
generators exists. Of course the answer is well known:
a Cartan-Weyl basis of generators exists iff the Lie
algebra is semisimple \cite{jacobson}.
In the next subsections we will outline the main ideas of the proof.
This exercise will also serve to help us finding the appropriate
notion of semisimplicity for Lie 3 (or higher $n$)-algebras 
in order for something like the Cartan-Weyl
basis to exist. 

\subsection{Properties of root space for 
metric Lie algebras}

By definition, a metric on a Lie algebra is a bilinear symmetric form which 
is  invariant. A Killing metric can always be constructed on any Lie algebra.
In general, it is
possible to have more than one metric on a Lie algebra. 
On simple Lie algebras, the Killing metric becomes the unique (up to 
a nonvanishing scaling factor) metric and is non-degenerate
\footnote{A metric is non-degenerate if
$\la x,y \ra =0$ for all $y$ implies that $x=0$.
}.  

In the main text of this paper, we are interested in the root space
structure of a metric Lie $3$-algebra. Therefore it is
instructive to consider here first the more general case of a metric
Lie algebra before restricting to the case of a semisimple Lie algebra. 
In particular this means we do not  assume the form \eq{kroot}
of the metric, which holds only for the Killing metric.

Let $\mg$ be a Lie algebra, $\mh$ be a Cartan subalgebra and 
$\langle \cdot,\cdot
\rangle$ be a non-degenerate  metric. Then the following
lemmas \ref{lem1} - \ref{lem5} hold. 
The proofs of these statements can be found in 
\cite{jacobson}. We note that they can be modified straightforwardly to
work for any
Lie algebra with a non-degenerate  metric, not just with the Killing
metric. 
\begin{lem} 
\label{lem1}
If $\a, \b$ are any two weights and $\a+\b \neq 0$, then
$\mg^\a \perp \mg^\b$ relative to the metric. 
\end{lem}
\begin{lem} 
\label{lem2}
The metric is non-degenerate when restricted to $\mh$.
If $\a$ is a root, then $-\a$ is also a root. 
Moreover $\mg^\a$ and $\mg^{-\a}$ are dual spaces relative to the metric.
\end{lem}
\begin{cor}
The metric is non-degenerate when restricted to $\mh$. 
\end{cor}
As a result of the corollary, we have that 
for any $\r\in \mh^*$, there exists a unique
$h_\r\in \mh$ such that
\be \label{h_rho}
\r(h) = \langle h_\r, h \rangle, \quad \mbox{for all $h \in \mh$}. 
\ee
The mapping $\r \to h_\r$ is 1-1 and subjective. Moreover, \eq{h_rho}
induces   a non-degenerate bilinear form on $\mh^*$:
\be \label{rho-sigma}
\langle \r,\s \rangle := \langle h_\r, h_\s \rangle , \quad
\mbox{for $\r,\s \in \mh^*$}.
\ee
It follows immediately that:
\begin{lem}\label{lem4}
Let $e_\a \in \mg^\a$ be such that $[h,e_\a] = \a(h) e_\a$ 
for all $h\in \mh^*$, then it is
\be
[e_\a,e_{-\a}] = \langle e_\a,e_{-\a} \rangle h_\a,
\ee
for all $e_{-\a} \in \mg^{-\a}$. 
Here $h_\a$ is defined as in \eq{h_rho}. Moreover due to \ref{lem2},
one can choose $\langle e_\a,e_{-\a} \rangle =1$.
\end{lem}
\begin{proof}
By the invariance of the metric, it is 
$\langle [e_\a,e_{-\a}], h \rangle =  \langle e_{-\a}, [e_\a,h] \rangle
= \langle e_{-\a}, e_\a \rangle \a(h)$. On the hand, \eq{h_rho} implies that
$\langle \langle e_\a,e_{-\a}\rangle h_\a , h \rangle 
= \langle e_{-\a}, e_\a \rangle \langle h_\a, h \rangle 
= \langle e_{-\a}, e_\a \rangle \a(h)$. The claim follows from the
non-degeneracy of the metric $\langle \cdot, \cdot \rangle$ when 
restricted to $\mh$. 
\end{proof}
\begin{lem}
\label{lem5}
If $\a$ is a root with nonvanishing norm, then
$n_\a = {\rm dim} \mg^\a =1$. Moreover the only integral multiples $k
\a$ of $\a$ which are roots are $\a, 0, -\a$. 
\end{lem}
\begin{proof}
For instruction, we outline the proof in \cite{jacobson}
here. First of all, since the metric on $\mh$ is non-degenerate, it
follows immediately from \eq{rho-sigma} that $\la \a,\a \ra \neq 0$.
Next consider $e_\a, e_{-\a}$ and $h_\a$ as defined in lemma
\ref{lem4} and set
\be
\cR = \CC e_\a \oplus \CC h_\a \oplus \sum_{k=1}^\infty \mg^{-k \a},
\ee
then $\cR$ is an invariant subspace of $\mg$ relative to ${\rm ad} h,
h \in \mh$ since
\be
[h,e_\a] = \a(h) e_\a, \quad [h, h_\a] =0, \quad 
[h, \mg^{-k \a}] \subset \mg^{-k\a}. 
\ee
The restriction of ${\rm ad} h$ to $\mg^{-k \a}$ has the single
characteristic root $-k \a(h)$. Hence we have
$\tr_{\cR} ({\rm ad} \,h) = \a(h) [ 1- n_{-\a} -2 n_{-2\a} -\cdots ]$.
In particular
\be \label{nnn}
\tr_{\cR} ({\rm ad} \,h_\a ) =\langle \a,\a \rangle 
\big[ 1- n_{-\a} -2 n_{-2\a} -\cdots \big].
\ee
Since $\cR$ is a subalgebra containing $e_\a$ and $e_{-\a}$, it is
invariant under ${\rm ad}\, e_\a$ and ${\rm ad}\, e_{-\a}$; and since
$h_\a=  [e_\a,e_{-\a}]$, therefore $[{\rm ad}\, e_\a,{\rm ad}\,
e_{-\a} ] =  {\rm ad}\, h_\a$ and hence $\tr_{\cR} ({\rm ad} \,h_\a )
=0$. The equation \eq{nnn} has the only solution $n_{-\a} =1, n_{-2\a}
= \cdots = 0$. Replacing $\a$ by $-\a$, we obtain our claim.
\end{proof}
As a result of this lemma, we obtain 
\begin{thm} \label{decomp5}
Let $\mg$ be a Lie algebra
carrying a non-degenerate  metric and $\mh$ is a Cartan subalgebra, 
we have the root decomposition for $\mg$ 
\be  \label{rrr}
\mg = \underset{\underbrace{}_{\mbox{Non-abelian}}}{\qquad\mh\qquad}
\oplus 
\underbrace{\Bigg(\bigoplus_{\langle \a,\a \rangle \neq 0} \mg^{\a}
\Bigg)}_{\mbox{dimension 1}}
\oplus 
\underbrace{\Bigg(\bigoplus_{\langle \a,\a \rangle = 0} \mg^{\a} 
\Bigg)}_{\mbox{dimension arb.}} ,
\ee
where each of the root space $\mg^\a$ in the last factor is of
dimension one. In general nothing can be said about the dimensions of
the root space $\mg^\a$ with zero norm roots, i.e. 
$\langle \a,\a\rangle =0$. 
\end{thm}

\subsection{Properties of root space for 
semisimple Lie algebras}

For a semisimple algebra, the Killing
metric is not just non-degenerate, but also satisfies the important 
property \eq{kroot}.
Using this one can establish the followings:
\begin{lem}
\label{lem7}
For a semisimple Lie algebra $\mg$ with Cartan subalgebra $\mh$, it is
\be \label{ahh}
\a([\mh,\mh]) =0, \quad \mbox{for all $\a \in \mh^*$}.
\ee
\end{lem}
\begin{proof}
To see how \eq{kroot} can be put into use, let us show the proof of this lemma.
We have seen above that for any $\a \in \mh^*$, 
there exists a $h_\a \in \mh$ such that 
$\a(h) = \langle h_\a, h \rangle $ where $\langle \cdot,\cdot \rangle$ is 
the Killing metric. Therefore for $h=[h_1,h_2]$, it is 
$\a(h) = \tr \big( [ {\rm ad} \, h_1, {\rm ad} \, h_2 ] \; 
{\rm ad} \, h_\a \big) = 0$ since the matrices ${\rm ad} \, h_1$, 
${\rm ad} \, h_2$, ${\rm ad} \, h_\a$ are all upper triangular as $\mh$
is nilpotent and it is $\tr(ABC) =\tr(ACB)$ for any upper  
matrices $A,B,C$. 
\end{proof}

\begin{prop}
\label{lem8}
For a semisimple Lie algebra $\mg$, the Cartan subalgebra $\mh$ is Abelian.  
\end{prop}
\begin{proof}
Using again \eq{kroot} and \eq{ahh}, we have $\langle h',k \rangle =0$
for all $h' \in [\mh,\mh]$,  $k\in \mh$. Now since the Killing metric is
non-degenerate when restricted to $\mh$, therefore it must be $h'=0$. 
\end{proof}

\begin{prop}
\label{lem6}
For a semisimple Lie algebra, it is 
\be
\langle \a, \a \rangle \neq 0
\ee
for every nonzero roots $\a$ 
relative to the Killing metric $\langle \cdot,
\cdot \rangle$ on $\mh^*$.
\end{prop} 
\begin{proof}
We refer the readers to  section 4.1 of
\cite{jacobson} for the proof of this proposition.
\end{proof}

As a result, for semisimple Lie algebras,
the factor of root spaces with zero norm roots is absent. Moreover, 
the Cartan subalgebra is Abelian. 
This establishes the
existence of a  Cartan-Weyl basis \eq{cw2} for  semisimple Lie
algebras.
Denoting the generator of each root space component $\mg^\a$ by $E^\a$ and 
the generators of the Cartan subalgebra by $H_I$. 
Suitably normalizing the generators, the Killing metric  reads
\be \label{metric-cA}
\la E^\a, E^\b \ra = \d_{\a+\b}, \quad \la E^\a, H^I \ra =0, 
\quad g_{IJ} := \langle H_I,H_J \rangle,
\ee
where  $g_{IJ}$, the restriction of the Killing metric on the Cartan 
subalgebra, is non-degenerate for a semisimple Lie algebra.
The Lie brackets written in the Cartan-Weyl basis takes the form:
\bea \label{cw2-1}
[H_I,H_J] &=& 0, \nn\\
\;  [ H_I, E^\a ] &=& \a_I E^\a, \nn\\
\; [E^\a, E^\b] &=& \begin{cases}  
0, & \mbox{if $\a+\b \neq 0$ not  a root},\\
c(\a,\b) E^{\a+ \b}, & \mbox{if $\a+\b \neq 0$ is a root},\\
- \a \cdot H & \mbox{if $\a+\b =0$},
\end{cases} 
\eea
where, $\a \cdot H = \a_I g^{IJ} H_J$ and $g^{IJ}$ is the inverse of
$g_{IJ}$. Here we have used the invariance of the metric in 
deriving the relation for $[E^\a,E^{-\a}]$.


The coefficient $c(\a,\b)$ is antisymmetric in its arguments and
satisfies the following conditions:
\be \label{c1}
c(\a,\b) c(\g, \a+\b) + c(\b,\g) c(\a, \b+\g) +c(\g,\a) c(\b,
\g+\a) =0
\ee
if each of $\a+\b, \b+\g, \g+\a$ and $\a+\b+\g$ is a root; and
\be
c(\a,\b) c(\a+ \b,-\a) - c(-\a,\b) c(-\a+\b,\a) = \a\cdot \b
\ee
if $\a,\b$ are roots.
These conditions follow from the Jacobi identities  $[
[E^\a,E^\b], E^\g]] = \cdots $ and   
$[[E^\a,E^\b], E^{-\a}]] = \cdots$. The second condition can also be
written as
\be \label{c2}
c(\a,\b) c(-\a,-\b) - c(\a,-\b) c(-\a,\b) = \a\cdot \b
\ee
since $c(\a,\b) = c(\b,\g) =c(\g,\a)$ for roots such that $\a+\b+\g=0$. 
 
The equations \eq{c1} and \eq{c2} constraint the root
system and the coefficients $c(\a,\b)$ algebraically. Solving these
conditions gives a complete classification 
of semisimple Lie algebras \cite{jacobson}.

\section{Theory of Lie $n$-Algebras and Root Space 
Decomposition}

\subsection{Basic notions of semisimplicity} 

Let $\cA$ be a Lie $n$-algebra \cite{Fil,K1}.
An subspace $I \subset \cA$ is an {\it ideal} 
if $[I\cA \cdots \cA] \subset I$. It is an {\it subalgebra} if 
$[I I \cdots I] \subset
I$. A Lie $n$-algebra is said to be {\it simple}
if there is no proper ideal and dim $\cA >1$. Simple Lie $n$-algebras
have been classified and have a very simple structure.
\begin{thm} \label{str-simple}
A simple Lie $n$-algebra is isomorphic to one of the Lie $n$-algebra
$\cA_{p,q}$. The Lie $n$-algebra 
$\cA_{p,q}$ is a metric Lie $n$-algebra with signature $(p,q)$, $p+q
=n+1$. 
It has $n+1$ generators $e_i$, $i =1, \cdots, n+1$ and is defined
by  the metric
\be
\langle e_i, e_j \rangle = \ve_i \d_{ij}
\ee
and the $n$-bracket relations 
\be
[e_1, \cdots, \hat{e_i}, \cdots, e_{n+1}] = (-1)^i \ve_i e_i.
\ee 
The signs $\ve_i$ are
given by $(+ \cdots +)$ for $\cA_{0,n+1} := \cA_{n+1}$, 
$(-+ \cdots ++)$ for $\cA_{1,n}$,   $(--+\cdots +)$ for $\cA_{2,n-1}$ etc.
\end{thm}

To discuss semisimplicity, one needs the
concept of nilpotency. 
However as discussed by Kasymov \cite{K1}, there is a
spectrum of notions of nilpotency and solvability for Lie $n$-algebras.  
\begin{defn}
An ideal $I$ of an Lie $n$-algebra is called {\it $k$-nilpotent} if $I_k^r =0$
for some $r\geq 0$. 
An ideal $I$ of an Lie $n$-algebra is called {\it $k$-solvable} 
if $I_k^{(r)} =0$
for some $r\geq 0$.  
Here $I_k^r, I_k^{(r)}$ are defined inductively by: 
\bea
I_k^0 &:=& I,\quad I_k^{s+1} := 
\big[I_k^s,\underbrace{I,\cdots, I}_{k-1} \; ,\, 
\underbrace{\cA, \cdots, \cA}_{n-k}\big], \\
I_k^{(0)}&:=&I,\quad I_k^{(s+1)} := 
\big[\underbrace{I_k^{(s)},I_k^{(s)},\cdots, I_k^{(s)}}_k\; , \, 
\underbrace{\cA, \cdots, \cA}_{n-k}\big]
\eea
and  $k =2, \cdots, n$.
\end{defn}
The concept of $k$-solvability (or $k$-nilpotency) 
for different $k$ are distinct. 
Special cases are $k=n$ which is the one considered originally by
Filippov \cite{Fil} and $k=2$ which is considered originally by 
Kuzmin \cite{K1}.
It is clear that a $k$-solvable (respectively $k$-nilpotent) 
ideal $I$ is also   
$(k+1)$-solvable (respectively $(k+1)$-nilpotent). 

\begin{defn}
The largest $k$-solvable ideal of a finite dimensional Lie $n$-algebra
$\cA$ is called the {\it $k$-radical} $\mR_k(\cA)$.  A Lie $n$-algebra $\cA$
is called {\it $k$-semisimple}  if $\mR_k(\cA) =0$.
\end{defn}
It is obvious that if a  Lie $n$-algebra is $k$-semisimple, then it is
also $(k-1)$-semisimple. In particular, for $n=3$, a Lie
3-algebra which is
semisimple in the sense of Filippov is also semisimple in the
sense of Kuzmin. Ling \cite{ling} has shown that a  Lie
$n$-algebra semisimple in the Filippov sense can be written as a
direct sum of its simple ideals. Therefore Lie 3-algebras which are 
Filippov-semisimple are too restrictive to be useful in the BLG models. 

What about Kuzmin-semisimple Lie 3-algebras? It is easy to establish the 
followings:
\begin{prop}
Consider the Lorentzian algebra \cite{lor}
$\cA= \mg \oplus \CC(u,v)$, where $\mg$ is a Lie algebra.
$\cA$ has a metric which is given by an extension of the Killing
metric on $\mg$ to $\cA$ with the nonvanishing component 
$\langle u,v \rangle =1$, and has the 3-brackets 
\be \label{lor}
[u, g_1, g_2] = [g_1,g_2]_\mg, \quad 
[g_1,g_2,g_3] = -  \langle [g_1,g_2] , g_3\rangle_\mg v.
\ee
Then $\cA$
is solvable in the sense of Filippov. It 
is Kuzmin-semisimple (Kuzmin-solvable) iff the Lie algebra 
$\mg$ is semisimple (solvable). 
\end{prop}
In the same way, it is easy to show that:
\begin{prop}
The Lie 3-algebra with a maximally isotropic center \cite{Jind2,Jindn}
is Kuzmin-semisimple iff the Lie algebra factors present there are
semisimple.  
\end{prop}
Now it is known that BLG theories based on metric Lie
3-algebras with  maximally isotropic center describes gauge theories
on D2-branes. However there may still be other type of
Kuzmin-semisimple Lie 3-algebras which is more appropriate for the
description of multiple M2-branes.  We
will investigate this question below and propose a 
more specialised type of Kuzmin-semisimple  Lie 3-algebras that is suitable 
for this purpose.

\subsection{Root space decomposition}
 
One of the
powerful results about the structure of Lie algebras is the root space
decomposition of Lie algebras with respect to a Cartan subalgebra.  
It is natural to ask if a similar result holds for Lie
$n$-algebra. Let us examine it.

Let $\cA$ be a Lie $n$-algebra. A vector space $V$ is called a Lie
$n$ $\cA$-module if one can define a Lie $n$-algebra structure on the
direct sum $\cB:=V+\cA$ such that
\be
[V, \cA, \cdots, \cA]  \subset V.
\ee
In this case one can associate to any set of elements $(a)=(a_1,
\cdots, a_{n-1}) \in \cA^{\wedge (n-1)}$ a linear transformation
$\r_{(a)} = \r_{(a_1, \cdots, a_{n-1})}$ acting on $V$ in accordance
with the rule:
\be
\rho_{(a)} : v \longmapsto [v, a_1, \cdots, a_{n-1}].
\ee
The operators $\rho_{(a)}$ satisfy the relations
\be\label{rho1}
[\rho_{(a)},\rho_{(b)}] = \sum_{i=1}^{n-1} 
\rho_{(a_1,\cdots, a_i R_{(b)}, \cdots, a_{(n-1)})},
\ee
\be\label{rho2}
\rho_{( [a_1, \cdots, a_n], b_2, \cdots , b_{n-1})} =
\sum_{i=1}^{n-1} \r_{(a_i, b_2, \cdots, b_{n-1})} \rho_{(a_1,\cdots,
\hat{a_i},\cdots, a_n)},
\ee
where $R_{(a)}$ is the right multiplication 
$c R_{(a)}  = [c, a_1, \cdots, a_{n-1}]$ . In this case we say there is
a representation $\r$ of the Lie $n$-algebra $\cA$ on the space $V$. 
Note that the operators $\r_{(a)}$ form a Lie algebra. We will denote by
$L_\r(\cA)$ the Lie algebra
generated by the operators $\r_{(a)}$.
Note also that the operators $R_{(a)}$'s satisfy the same
relations \eq{rho1}, \eq{rho2}, so it form a representation of $\cA$ on 
$\cA$. This is called the {\it regular representation}. 
Using the regular representation $R$, 
one can introduce a {\it Killing form} for any Lie $n$-algebra by
\be
\k((a);(b)) := \tr( R_{(a)} R_{(b)})
\ee
where $(a) = (a_1, \cdots a_{n-1}), (b) = (b_1, \cdots b_{n-1})$.
The Killing form is invariant. Moreover it is known that \cite{K1,K2}: 
\begin{lem}
Let $\cA$ be  a Lie $n$-algebra, then 
\begin{itemize}
\item  (Cartan criterion of semisimplicity) 
$\cA$ is semisimple in the Kuzmin sense 
iff the Killing form is non-degenerate,
\item (Cartan criterion of solvability)
$\cA$ is solvable in the Kuzmin sense 
iff \\$\k(c_1, a_2, \cdots, a_{n-1}; c_2, b_2,
\cdots, b_{n-2}) =0$ for all $c_1, c_2 \in \cA^2$, $a_i, b_j \in \cA$.
\end{itemize}
\end{lem}

Let us denote the Lie algebra
generated by the operators $R_{(a)}$ as  $L(\cA)$.
Kasymov \cite{K1}
has shown that for a nilpotent (in the sense of Filippov) subalgebra
$\cH$ of the Lie $n$-algebra $\cA$,  the
subalgebra $L(\cH)$ generated by operators $R_{(h)}$, $(h) =(h_1,
\cdots, h_{n-1})$, $h_i \in \cH$ is a nilpotent subalgebra of the Lie algebra 
$L(\cA)$. 
Therefore the theorem \ref{decomp3-thm} for Lie algebra implies 
the following root space decomposition for a Lie
$n$-algebra:
\begin{thm} \label{n-decomp1}
Let $\cH$ be a nilpotent subalgebra (in the sense of Filippov) 
of a Lie $n$-algebra $\cA$, and $\r=R|_\cH$
is the regular representation of $\cH$ in $\cA$. Then we have the
decomposition
\be
\cA = \cA^0 \oplus \bigg( \bigoplus_{\a\in \D_{\cA}(\cH)} \cA^\a\bigg),
\ee
where
\bea \label{H0}
\cA^0 &=& \big\{ y \in \cA \,|\, y  R_{(H)}^m  =0 
\;\mbox{for some $m>0$ $\&$ for all $(H) \in \cH^{\wedge (n-1)}$} \big\},
\;\;\; \\
\cA^\a &=&  \big\{ y \in \cA \,|\, y (R_{(H)}- \a((H)) E)^m  =0 
\;\mbox{for some $m>0$ $\&$ for all $(H) \in \cH^{\wedge (n-1)}$} \big\}.
\nn
\eea 
Here $\Phi_\r(\cH) = \{\l_1, \cdots, \l_s \}$ is  collection of
the different weights $\l_i \in (\cH^{\wedge (n-1)})^*$ of the 
representation $\r$, and 
$\D_{\cA}(\cH) = \Phi_\r(\cH)/\{ 0 \}$ is the collection of all
nonzero weights (called roots).
\end{thm}
Note that roots are skew-symmetric in their argument and can be viewed 
as a $(n-1)$-forms, generalizing the concept of roots in ordinary Lie algebra.

In Lie algebra theory,  a Cartan subalgebra is defined as a subalgebra 
which is nilpotent and equal to its normalizer. 
Cartan subalgebras play a central role in 
the structure theory of finite dimensional Lie algebra. 
For example, as reviewed in section 2, 
the properties of Cartan subalgebras allow us to
classify finite dimensional semisimple Lie algebras completely.
  
With the different notions of nilpotency and solvability 
to choose from for a Lie $n$-algebras (for $n \geq 3$), 
there is also a possibility of different notions of Cartan subalgebras. It
turns out that the really useful definition is the one referring to
nilpotency in the Filippov sense. Defining the {\it normalizer} of a
subspace $V$  of $\cA$ to be the subspace $\mN(V)= \{
a \in \cA | [a, V, \cdots, V] \subset V \} $. We have the definition:
\begin{defn}
A subalgebra $\cH$ of a Lie $n$-algebra $\cA$ is a {\it Cartan subalgebra}
if it is nilpotent in the sense of Filippov and equal to its own
normalizer.  
\end{defn}
Obviously the dimension of a Cartan subalgebra of a Lie $n$-algebra is 
at least $n-1$.

With this definition, Kasymov  was able to
establish the existence of Cartan subalgebras for any 
finite dimensional Lie $n$-algebra $\cA$ over $\CC$
\cite{K1,K3}. 
Moreover all 
Cartan subalgebras are conjugate to each other \cite{K3}. 
For example, for the
Lorentzian 3-algebra \eq{lor}, one easily see that the subalgebra
defined by $\cH:= \{ u, v, h_i \}$ is a Cartan subalgebra. Here $h_i$
are the generators of the Cartan subalgebra of the Lie algebra $\mg$. 

Back to theorem \ref{n-decomp1}, one can show that  \cite{K1} $\cA^0 = \cH$
if $\cH$ is a Cartan subalgebra. As a result, 
theorem \ref{n-decomp1} is refined to:
\begin{thm}
Let $\cA$ be a Lie $n$-algebra and $\cH$ be a Cartan subalgebra of it,  
then we have the root space decomposition
\be \label{root-H}
\cA = \cH \oplus \bigg( \bigoplus_{\a\in \D_{\cA}(\cH)} \cA^\a\bigg).
\ee
\end{thm}
The root space components satisfy the $n$-bracket
\be \label{roots-CR}
[ \cA^{\a_1}, \cdots, \cA^{\a_n}] \begin{cases}
\subset \cA^{\a_1 + \cdots + \a_n}, &  \a_1+ \cdots + \a_n \in
\Phi_R(\cH) \\
=0 & \mbox{otherwise}
\end{cases}
\ee
Moreover, in analogous to Lie algebras,
the Killing form on $\cH^{(n-1)}$ can be expressed in terms of the roots
\be \label{kroot-2}
\k( (h); (h')) = \sum_\a n_\a \a((h)) \a((h')). 
\ee
This completes our review of the theory of Lie $n$-algebra known in the
literature.

\subsection{Properties of root spaces for metric 
Lie $n$-algebras} \label{rs}

Next let us investigate further the properties of this root
space decomposition for a Lie $n$-algebra when there is additionally a
non-degenerate  metric. 
Following essentially the same analysis as for 
Lie algebras, it is not difficult to show that:
\begin{lem} \label{lem3a}
Let $\cA$ be a Lie $n$-algebra and $\langle \cdot,\cdot \rangle$  a
non-degenerate  metric on $\cA$. Then,
(i) For any two weights $\a, \b$ 
and $\a +\b \neq 0$, $\cA^\a \perp \cA^\b$
relative to the metric. (ii) The metric is non-degenerate when 
restricted to a Cartan subalgebra
$\cH$. 
(iii) If $\a$ is a root, then -$\a$ is
also a root and $\cA^\a$ and $\cA^{-\a}$ are dual spaces relative to the
metric. 
\end{lem}
One can be quite explicit with the metric. 
If one denotes the generators of the Cartan subalgebra $\cH$ by 
$H_I$ and the 
generators of the root space components $\cA^\a$ by $E^\a_i$, 
$i=1,\cdots, n_\a$ where $n_\a = {\rm dim} \cA^\a$, i.e.
\be \label{EH-basis} 
\{T^A\} =\{H_I,E^\a_i\}, \qquad \mbox{where} \quad 
I=1, \cdots, N, \quad \mbox{and}\quad i=1,\cdots, n_\a,
\ee
then one can always choose the basis of generators such 
that the metric takes the form
\be \label{EH-metric}
\la H_I, H_J \ra =g_{IJ}, \quad \la H_I, E^\a_i \ra =0, \quad
\la   E^\a_i ,  E^\b_j \ra = \d_{ij} \d^{\a+\b},
\ee
where $g_{IJ}$ is non-degenerate.  To obtain a real Lie $n$-algebra, 
one can impose the Hermitian structure
\be \label{Hermitian}
(H_I)^\dagger = H_I, \quad (E^\a_i)^\dagger = E^{-\a}_i.
\ee
In this case, the index $m$ is the number of negative eigenvalues of the 
metric $g_{IJ}$.

With the non-degenerate metric, one can associate to each root $\rho$
and fixed $h_1, \cdots$, $h_{n-2} \in \cH$, a unique element $h_{(\rho; h_1,
\cdots, h_{n-2})} \in \cH$ such that
\be\label{rhoh}
\rho(h_1, \cdots, h_{n-2}, h) = \la h_{(\rho; h_1,
\cdots, h_{n-2})}, h \ra, \quad \mbox{for all $h \in \cH$}.
\ee 
It is easy to establish  a statement similar to the lemma \ref{lem4} for 
Lie algebras:
\begin{lem} \label{lem36}
Let $e_\a \in \cA^\a$ be such that 
$[e_\a, h'_1, \cdots, h'_{n-1}] = \a((h'))
e_\a$ for all $(h') =(h'_1, \cdots, h'_{n-1})$, 
$h'_i \in \cH$, and let
$e_{ -\a} \in \cA^{-\a}$, then
\be \label{ha-def}
[e_\a, e_{-\a}, h_1, \cdots, h_{n-2}] = \la e_{-\a}, e_\a\ra  
h_{(\a; h_1, \cdots, h_{n-2})} (-1)^n
\ee
for $h_1, \cdots, h_{n-2} \in \cH$.
\end{lem}
At this point, one may then try to establish something 
like lemma \ref{lem5} about the
dimension of the root space components $\cA^\a$. But  
one sees an essential obstacle immediately. For Lie algebras, we have
the relation $[h,h_\a]=0$ for all $h \in \mh$. 
This follows from the definition of $h_\a$ 
and  the Jacobi identity.  This relation has played an essential role 
in establishing the lemma \ref{lem5}. 
For Lie $n$-algebra, by using the definition \eq{ha-def} and the fundamental 
identity, one can see that the relation
$[h'_1, \cdots, h'_{n-1}, h_{(\a; h_1, \cdots, h_{n-2})}] =0$  for
arbitrary $h'_i, h_j \in \cH$ is true only if $\cH$ is Abelian. Let us
therefore ask when is $\cH$  Abelian. 

We recall that in the case of a Lie 
algebra with a metric being non-degenerate on the Cartan subalgebra,  
one can show immediately that 
the Cartan subalgebra is Abelian if the condition $\a([\mh,\mh])
=0$ holds (see proposition \ref{lem8}). This is so, for example, if the Lie
algebra is semisimple (Lemma \ref{lem7}). 
For higher Lie $n$-algebra, one can similarly
show the following:
\begin{lem} \label{lem2cond}
Let $\cA$ be a Lie $n$-algebra and $\cH$ be a Cartan subalgebra. If 
\begin{itemize}
\item[i.]
the Killing form is non-degenerate on the subspace generated by $(a_1,
\cdots, a_{n-1})$, where $a_i \in  \cH$,

\item[ii.] and
\be \label{ahhh}
\a( [h_1, \cdots, h_n], h'_2, \cdots , h'_{n-1}) =0, \quad
\mbox{for all $h_i, h'_j \in \cH$},
\ee
\end{itemize}
then $\cH$ is Abelian.
\end{lem}
\begin{proof} 
Using \eq{kroot-2}, we have 
$\k([h_1, \cdots, h_n], \tilde{h}_2, \cdots , \tilde{h}_{n-1}; 
\tilde{\tilde{h}}_1, \cdots, \tilde{\tilde{h}}_{n-1}) =0$ 
for arbitrary $h_i, \tilde{h}_j,\tilde{\tilde{h}} _k \in \cH$. 
The stated non-degeneracy of 
the Killing form thus implies that  $[h_1, \cdots, h_n]=0$.
\end{proof}

Let us examine these two conditions more closely and ask when do these
two conditions hold.
For the first condition, we recall
that the Killing form  is non-degenerate iff the  
Lie $n$-algebra is semisimple in the Kuzmin sense. However even in
this case, the non-degeneracy of the Killing form only implies (using
its invariance)
that the Killing form is non-degenerate when 
restricted on the subspace generated by $(a_1,
\cdots, a_{n-1})$, 
where $a_i \in \cA^{\l_i}$ and $\l_1 + \cdots + \l_{n-1} =0$ . 
This is  
weaker than the first condition except for the case of $n=2$ where  
the non-degeneracy of the Killing metric of a Lie algebra implies that it is 
also non-degenerate when restricted on a Cartan subalgebra.  
As for the second condition, we recall that for Lie algebras, the condition
$\a([\mh,\mh]) =0$ is a consequence of the fact that for each $\a \in \mh^*$,
there exists an unique $h_\a \in \mh^*$ such that 
\be \label{hah}
\a(h) = \langle h_\a, h\rangle
\ee 
for all $h\in \mh$
\footnote{
As a result,
$\a([h_1,h_2]) = \langle h_\a, [h_1,h_2] \rangle 
= \tr ({\rm ad}\,{[h_1,h_2]} {\rm ad}\, h) 
= \tr ([{\rm ad}\, h_1, {\rm ad}\, h_2] {\rm ad}\, h) =0$
since ${\rm ad}\, {h_1}$, ${\rm ad}\, {h_2}$ and  ${\rm ad}\, {h}$ are 
nilpotent matrices.
}.  
For Lie $n$-algebras, the natural
generalization of \eq{hah} is to require that
for each $\a \in (\cH^{\wedge (n-1)})^*$, 
there exists  a unique $(h_\a) \in \cH^{\wedge (n-1)}$
such that
\be \label{akappa}
\a((h)) = \k((h); (h_\a)),
\ee
where $\k$ is the Killing form. Assuming its validity, we have therefore
\bea \label{a1234}
\a([h_1, \cdots, h_n], h'_2, \cdots , h'_{n-1}) &=&
\tr( R_{([h_1, \cdots, h_n], h'_2, \cdots , h'_{n-1})} R_{(h_\a)} ).
\eea
However this is nonvanishing since even though 
\be
R_{([h_1,\cdots, h_n] , h_2',\cdots, h'_{n-1})} =
\sum_{i=1}^n (-1)^{i-1} 
R_{(h_i,h'_1, \cdots, h'_{n-1})} R_{(h_1, \cdots, \hat{h}_i,
\cdots,h_i)},
\ee
and the order of $R$ in the trace does not matter (this is because 
all the $R$'s 
involved are nilpotent matrices and one has $\tr (abc) = \tr(acb)$
for nilpotent matrices $a,b,c$),  there is no complete cancellation as
in the  case for Lie algebras.
Thus we conclude the condition \eq{ahhh} does not follow from \eq{akappa}.
Instead, the condition \eq{ahhh} appears to be an 
independent condition that is of
interest to be studied for its own right.

As a result, we conclude that the two conditions listed in the lemma
\ref{lem2cond} do not follow from 
$k$-semisimplicity.
In general, for a $k$-semisimple metric Lie $n$-algebra $\cA$ ($k < n$), 
\begin{itemize}
\item[1.] the Cartan subalgebra is non-abelian, and 
\item[2.] the root space components $\cA^\a$ are generally of
dimension greater than 1. 
\end{itemize}
i.e. the root space decomposition for a  $k$-semisimple Lie
$n$-algebra  takes the form
\be \label{decomp-non}
\cA =  \underset{\underbrace{}_{\mbox{Non-abelian}}}{\qquad\cH\qquad}
 \oplus\underbrace{\bigg( \bigoplus_{\a\in \D_{\cA}(\cH)} 
\cA^\a
\bigg)}_{\mbox{dimension arb.}}.
\ee
This is much more complicated than that of a semisimple 
Lie algebra.

\section{Strong-Semisimplicity and Generalized 
Cartan-Weyl $n$-Algebras}

When one of the worldvolume spatial dimension is compactified, 
a system of multiple M2-branes
reduces to a system of multiple D2-branes whose worldvolume theory 
is equipped with a $U(N)$ gauge symmetry. Therefore 
in order for a BLG theory to describe multiple M2-branes, 
there should be a connection between the
3-algebra describing the symmetry of the multiple M2-branes and the 
(semisimple) Lie algebra describing the symmetry of multiple D-branes 
systems.

\subsection{A reduction condition: strong-semisimplicity}

Let us start with a simple observation about  general Lie $n$-algebras. 
Consider a Lie $n$-algebra $\cA$.  Let us fix $\hb_1, \cdots, \hb_{n-2} \in
\cA$ and consider a 2-bracket defined by
\be 
[x,y]_{\la \hb \ra}:= [x,y, \hb_1, \cdots, \hb_{n-2}].
\ee 
Here we have used the symbol $\la \hb \ra$ to denote the collection of the
$(n-2)$ elements $\hb_1, \cdots, \hb_{n-2}$.
It is easy to see that the 2-bracket $[\cdot,
\cdot]_{\la \hb \ra}$ obeys the Jacobi identity as a result of the fundamental
identity for the $n$-bracket. This allows one to introduce a 
Lie algebra structure on any Lie $n$-algebra. Let us denote the
corresponding Lie algebra 
as $\cA_{\la \hb \ra}:= (\cA, [\cdot,\cdot]_{\la \hb \ra})$.
More generally, one can 
fix a collection of $p$ elements 
$\hb_1, \cdots, \hb_{p} \in \cA$, $1 \leq p \leq n-2$, 
and consider the $(n-p)$-bracket
\be
[x_1,\cdots, x_{n-p}]_{ (\hb_1, \cdots, \hb_{p})}
:= [x_1,\cdots, x_{n-p}, \hb_1, \cdots, \hb_{p}],
\quad x_i \in \cA.
\ee
The algebra $\cA_{(\hb_1,\cdots, \hb_{p})} 
:= (\cA, [\cdot,\cdots, \cdot]_{(\hb_1,\cdots, \hb_{p})})$ form a 
Lie $(n-p)$-algebra.

Let us
denote the $k$-radical of the Lie $(n-p)$-algebra $\cA_{(\hb_1,\cdots,
\hb_{p})}$ by $\mR_{k,(\hb_1,\cdots,\hb_{p})}(\cA)$. Obviously the
$k$-radical is defined only for $k\leq n-p$.
\begin{defn}
Let $\cA$ be a  Lie $n$-algebra $\cA$ and $2 \leq k \leq n$ fixed. We
say that $\cA$ 
is {\it $(k,p)$-semisimple}, $0 \leq p\leq n-k$, 
if there exists $\hb_1, \cdots , \hb_p \in \cA$ such that
$\mR_{k,(\hb_1, \cdots, \hb_p)} =0$, i.e. if the associated Lie
$(n-p)$-algebra $\cA_{(\hb_1,\cdots,\hb_p)}$ is semisimple. 
\end{defn}
It is easy to establish that:
\begin{lem} 
Let $\cA$ be a Lie $n$-algebra and $2 \leq k \leq n$ fixed. 
Take $n-k$ elements $\hb_1, \cdots \hb_{n-k} \in \cA$. For each $0
\leq p \leq n-k$, the $k$-radicals of the associated Lie $(n-p)$-algebras 
$\cA_{(\hb_1, \cdots, \hb_p)}$  satisfy
\be
\mR_k (\cA) \subset \mR_{k,(\hb_1)} (\cA) \subset \cdots \subset \mR_{k,
(\hb_1, \cdots, \hb_{n-k})} (\cA).
\ee
\end{lem}
\begin{proof}
Denote $I= \mR_{k,(\hb_1,\cdots, \hb_p)}$ for some fixed $p$. The fact that $I$ is 
a maximal $k$-solvable ideal of $\cA$ implies that $I^{(r)}_k=0$ for some $r$
where we recall the definition of  $I^{(r)}_k$: 
\be
I^{(r)}_k:= [\underbrace{I_k^{(r-1)}, \cdots, I_k^{(r-1)}}_k, 
\underbrace{\cA, \cdots, \cA}_{n-p-k}, \underbrace{\hb_1, \cdots,\hb_p}_p] .
\ee
By restricting one of the $\cA$ to be $\hb_{p+1}$, we obtain
\be
 [\underbrace{I_k^{(r-1)}, \cdots, I_k^{(r-1)}}_k, 
\underbrace{\cA, \cdots, \cA}_{n-p-k-1}, 
\underbrace{\hb_{p+1}, \hb_1, \cdots,\hb_p}_{p+1}] =0.
\ee
This implies that $I$ is a $k$-solvable ideal of the Lie $(n-p-1)$-algebra
$\cA_{(\hb_1, \cdots, \hb_p,\hb_{p+1})}$ and so
$I \subset \mR_{k,(\hb_1, \cdots, \hb_p, \hb_{p+1})} =0$. Hence the claim.
\end{proof}
As a result of this lemma, it is clear that
$(k,p)$-semisimplicity implies $(k, p-1)$-semisimplicity. 
Therefore for fixed $k$, $(k,n-k)$-semisimplicity is the strongest and
$(k,0)$-semisimplicity is the weakest. Note that $(k,0)$-semisimplicity is
the $k$-semisimplicity originally introduced by
Kasymov, with $(2,0)$-semisimplicity being Kuzmin sense of
semisimplicity and $(n,0)$-semisimplicity being the Filippov sense  
of semisimplicity.
In view of this, one can introduce a graded notion of $k$-semisimplicity
(with $k$ fixed) for a Lie $n$-algebra and we have the following
web of $(k,p)$-semisimplicity for Lie $n$-algebras.  For a
fixed $ 2\leq K\leq n$, we have
\be
\begin{array}{cccccccccl}
(K,0)        & \Leftarrow  & (K,1)      & \cdots & \Leftarrow  
& (K,n-K) & &  &   &  \\
\Downarrow   &  & \Downarrow            &  &  &  \Downarrow  
&&  &   & \nn\\
(K-1,0)        & \Leftarrow &  (K-1,1)      & \cdots & \Leftarrow & 
(K-1,n-K) & \Leftarrow &  (K-1,n-K+1)&   &  \\
\vdots & & \vdots& & & \vdots   &  &\vdots&\ddots &\\
\Downarrow   &  & \Downarrow            &  &  &  \Downarrow  &&
\Downarrow &   & \\
(2,0) & \Leftarrow &  (2,1)    & \cdots & \Leftarrow &  (2,n-K) 
& \Leftarrow &  (2,n-K+1) & \cdots & \Leftarrow (2, n-2).
\end{array}
\ee
For example, for Lie 3-algebras and for $K=3$, we have the
following relationship among the various definitions of semisimplicity:
\be
\begin{array}{ccc}
\mbox{(3,0) = Filippov-semisimplicity }&&  \nn \\
\Downarrow && \nn\\
\mbox{(2,0) = Kuzmin-semisimplicity} &\Leftarrow& \mbox{(2,1)-semisimplicity}.
\end{array}
\ee 

We have seen that a metric Lie $n$-algebra which is
semisimple in the Kuzmin sense has a non-degenerate Killing form.
However, except for the case of $n=2$, 
this does not lead to a simple root space decomposition. We also know that 
the  more
restrictive notion of Filippov-semisimplicity
(i.e. $(n,0)$-semisimplicity)
is too strong as it
implies that the Lie $n$-algebra is a direct sum of simple factors like
$\cA_{q, n-q}$ and Abelian ones. What about the other notions of
semisimplicity? Does any of them imply a simple structure 
of the root space?
 
To proceed, we make the following observation:
Naively one may try to relate the Lie algebra $\cA_{\la \hb \ra}$ (with
$\hb_1, \cdots, \hb_{n-2} \in \cA$) with the Lie algebra living on the
D-branes obtained by compactification. However it is certainly too strong, and
also unreasonable to demand this to be true for any choice of the
elements $\hb_i$. It turns out, due to the following lemma, a
distinguished choice is to  take the $\hb_i$'s to 
be in a Cartan subalgebra $\cH$ of $\cA$. 
 
\begin{lem}
Let $\cA$ be a Lie $n$-algebra and $\cH$  a Cartan subalgebra. Then 
$\cH$ is also a  
Cartan subalgebra of the associated Lie $(n-p)$-algebra 
$\cA_{(\hb_1,\cdots,\hb_{p})}$
if  $\hb_1, \cdots, \hb_{p} \in \cH$. 
\end{lem} 
\begin{proof}
That $\cH$ is a nilpotent subalgebra of 
$\cA_{(\hb_1,\cdots,\hb_{p})}$ follows immediately from 
the fact that
$\cH$ is a nilpotent subalgebra (in the sense of Filippov) of $\cA$.
Next consider $x$ in the normalizer of 
$N_{(\hb_1,\cdots,\hb_{p})}(\cH)$ of $\cH$ in 
$\cA_{(\hb_1,\cdots,\hb_{p})}$, i.e. 
$[x, \underbrace{\cH, \cdots, \cH}_{n-p-1}, 
\hb_1, \cdots, \hb_p] \in \cH$.
Due to the property \eq{roots-CR} of the $n$-bracket, it follows immediately 
that $x \in \cH$ and hence $N_{(\hb_1,\cdots,\hb_{p})}(\cH) = \cH$.
Therefore 
$\cH$ is a Cartan subalgebra of $\cA_{(\hb_1,\cdots,\hb_{n-p})}$.
\end{proof}
\begin{defn} \label{sss}
Let $\cA$ be a Lie $n$-algebra and $\cH$  a Cartan subalgebra.
We say that $\cA$ is {\it strong-semisimple} if there exists
$\hb_1, \cdots, \hb_{n-2} \in \cH$ such that the Lie algebra
$\cA_{\la \hb \ra }$ is semisimple, i.e. $\cA$ is $(2,n-2)$-semisimple
with respect to a choice of elements $\hb_1, \cdots, \hb_{n-2} \in \cH$. 
\end{defn}

Since under suitable conditions, 
a system of multiple M2-branes can always be reduced to  
a system of multiple D2-branes, one expects a natural connection between 
the Lie 3-algebraic symmetry of the system of M2-branes and  
the semisimple Lie algebra of the system of D2-branes.
It seems therefore quite reasonable to require  that the Lie 3-algebra 
to be employed in the BLG theory  to be strong-semisimple.
We conjecture that 
strong-semisimple Lie $3$-algebras is the appropriate kind 
of Lie 3-algebras that is 
relevant for describing the symmetry of multiple M2-branes.

In the following, 
we will analysis
the structure of a strong-semisimple Lie $n$-algebra in more details.
In particular we will show that 
the notion of strong-semisimplicity leads to great simplification of the 
root space decomposition \eq{decomp-non}, yielding
a simple and yet rich structure for the root-space decomposition
of a Lie $n$-algebra, see theorem \ref{root-space-n} and explicitly
\eq{cw31s}-\eq{cw34cs}.

\subsection{Generalized Cartan-Weyl $n$-algebras} 

Consider a strong-semisimple Lie $n$-algebra $\cA$ and let $\cH$ be a 
Cartan subalgebra. 
Let $\hb_1, \cdots, \hb_{n-2} \in \cH$ be a choice of $n-2$ elements such that
the Lie algebra $\cA_{\la \hb \ra }$ is  semisimple. 
Using the fact that the Killing metric $\bar{\k}$ on $\cA_{\la \hb \ra }$
is non-degenerate \footnote{
It has a simple relation with the Killing form on the Lie $n$-algebra $\cA$
\be
\bar{\k} (x,y) = \k(x, \hb_1, \cdots, \hb_{n-2}; y, \hb_1, \cdots, \hb_{n-2}),
\quad x,y \in \cA.
\ee
}, 
one can establish the following important property for the
structure of the root space decomposition for a strong-semisimple Lie
$n$-algebra. 
\begin{thm} \label{root-space-n}
Let $\cA$ be a Lie $n$-algebra which is strong-semisimple, then 
the root space decomposition for $\cA$ takes the form
\be
\cA = \cH \oplus \underbrace{\bigg( \bigoplus_{\a\in \D_{\cA}(\cH)} 
\cA^\a \bigg)}_{\mbox{dimension 1}},
\ee
where $\cA^\a = \big\{ y \in \cA \,|\, [y, (h)]=  \a((h)) \; y, \;
\mbox{for all $(h) \in \cH^{\wedge (n-1)}$} \big\}$ 
is of dimension 1.  Moreover, the only
integral multiples of $\a$ which are roots are $0, \pm \a$.
\end{thm}
\begin{proof}
The proof is a slight generalization of the proof of lemma \ref{lem5}. 
Let $\hb_1, \cdots,$ $\hb_{n-2} \in \cH$ be the set of elements such that
$\cA_{(\hb_1, \cdots, \hb_{n-2})}$ is semisimple. Since the Killing metric
$\bar{\k}$ is non-degenerate, we can use it in \eq{rhoh} and lemma
\ref{lem36}. 
Therefore  
corresponds to each nonzero root $\a$ of the Lie $n$-algebra, 
one can associate an element $h_{(\a;\hb_1,\cdots,\hb_{n-2})}\in \cH$ such
that
\be \label{khh}
\bar{\k}(h_{(\a;\hb_1,\cdots,\hb_{n-2})}, h) = \a(\hb_1, \cdots, \hb_{n-2}, h).
\ee
Moreover as in lemma \ref{lem36} we have elements $e_\a$, $e_{-\a} \in
\cA$ such that 
\be \label{eehh}
[e_\a, e_{-\a}, \hb_1, \cdots, \hb_{n-2} ] = \bar{\k}(e_{-\a}, e_\a)
h_{(\a;\hb_1,\cdots,\hb_{n-2})} (-1)^n.
\ee 
Next, as in the proof of lemma \ref{lem5} we introduce the vector space 
\be
\cR :=\CC e_\a \oplus\CC h_{(\a;\hb_1,\cdots,\hb_{n-2})} \oplus
\sum_{k=1}^\infty \cA^{-k \a}.
\ee
It is 
\bea
\; && [e_\a, \hb_1, \cdots, \hb_{n-2}, h] 
= \a( \hb_1, \cdots, \hb_{n-2}, h) e_\a, 
\\ 
\; && [ h_{(\a;\hb_1,\cdots,\hb_{n-2})}, \hb_1, \cdots, \hb_{n-2}, h] =0, 
\quad [\cA^{-k\a}, \hb_1, \cdots, \hb_{n-2}, h] \subset \cA^{-k\a}, \;\;
\eea
for all $h \in \cH$. 
Therefore 
\be
\tr_{\cR} (R_{(\hb_1, \cdots, \hb_{n-2}, h)}) = \a(\hb_1, \cdots, \hb_{n-2}, h)
\big[
1-n_{-\a} - 2n_{-2 \a} - \cdots
\big].
\ee
Finally we notice that for $h =  h_{(\a;\hb_1,\cdots,\hb_{n-2})}$, we can
use \eq{eehh} to obtain that 
$R_{(\hb_1, \cdots, \hb_{n-2}, h)} \sim 
[R_{(e_\a, \hb_1, \cdots, \hb_{n-2})}), 
R_{(e_{-\a}, \hb_1, \cdots, \hb_{n-2})})]$ 
and hence 
its trace is zero. Moreover \eq{khh} implies that for this $h$, $\a(\hb_1,
\cdots \hb_{n-2}, h) = \bar{\k}(h,h) \neq 0$, 
therefore we obtain $n_{-\a} =1, n_{-2\a} = \cdots = 0$
and hence our claim.
\end{proof}

The theorem \ref{root-space-n} states that a
strong-semisimple Lie $n$-algebra has a basis of generators consisting of
a number of step generators $E^\a$ whose roots are non-degenerate and a 
number of generators $H_I$ which spans the Cartan subalgebra $\cH$.
The $n$-brackets expressed in terms of these generators take the form
\bea
\; [H_{I_1}, \cdots, H_{I_n}] &=& L_{I_1 \cdots I_n}{}^M H_M, \\
\; [ H_{I_1},\cdots, H_{I_{n-1}}, E^\a] &=& \a_{I_1 \cdots I_{n-1}} E^\a, 
\eea
together with the other relations that follow from \eq{roots-CR}. Here
$L_{I_1 \cdots I_n}{}^M$ is a set of constants such that $\cH$ form
a Cartan subalgebra. 
In general for $n\geq 3$, the Cartan subalgebra  is  non-abelian. 
The brackets are further constrained if there is an invariant metric. 
For the first nontrivial case of $n=3$, the 3-brackets of a
metric strong-semisimple Lie 3-algebra 
take the form:
\bea 
\; [H_I, H_J, H_K] &=&  L_{I J K}{}^M H_M, \label{cw31s} \\
\; [H_I, H_J, E^\a]& =& \a_{IJ} E^\a, \label{cw32s}
\eea
\begin{subnumcases}{[H_I,E^\a, E^\b] =}
\a_{IK} g^{KL} H_L, & \mbox{if $\a+ \b =0$,} \label{cw33as}\\
g_I(\a,\b) E^{\a+\b}, & \mbox{if $\a+ \b \neq 0$ is a root,} \label{cw33bs}\\
0,  & \mbox{if $\a+ \b$ is not a root,} \label{cw33cs}
\end{subnumcases}
\begin{subnumcases}{[E^\a,E^\b, E^\g] =}
- g_K(\a,\b) g^{KL} H_L,  & \mbox{if $\a+ \b + \g =0$,} \label{cw34as} \\
c(\a,\b,\g) E^{\a+\b+\g},  & \mbox{if $\a+ \b + \g \neq 0$ a root,} 
 \label{cw34bs}\\
0,  & \mbox{if $\a+ \b + \g$ is not a root.}   \label{cw34cs}
\end{subnumcases}

Within this framework, we see that  Cartan-Weyl $n$-algebras 
are just special form of  strong-semisimple metric Lie $n$-algebras in
the case that the 
Cartan subalgebra is Abelian.
Therefore to be uniform with terminology, 
we will call strong-semisimple metric Lie
$n$-algebras as {\it generalized  Cartan-Weyl $n$-algebras}.

Given its possible relevance for describing multiple M2-branes and
also due to its naturalness from  a pure mathematical point of view, 
it is an interesting question to classify the 
generalized  Cartan-Weyl $n$-algebras, particularly for the case
$n=3$.
To do this, one needs to solve for 
the consistency conditions similar to those 
performed for the
Cartan-Weyl 3-algebras where $L_{IJK}{}^M =0$ (see appendix A of
\cite{cat1}). 
This is a more difficult task and we will leave it for 
further investigation.
In the next subsection, we will consider  
a certain special class of generalized  Cartan-Weyl $3$-algebras
where it is  possible to obtain a rather 
explicit classification.

\subsection{A special class of generalized 
Cartan-Weyl $3$-algebras}
\label{ccw3}

If one examines carefully our
construction in the appendix A of \cite{cat1} for the  Cartan-Weyl 
3-algebras, 
one can see that we have indeed only used the following 
relations  in our  consistency checks,
\bea \label{H-cond1}
\; [[H_I,H_J,H_K], H_J , E^\a] &=& 0, \\
\; [[H_I,H_J,H_K], E^\a , E^\b]
&=& 0, \quad \mbox{for any $H$'s and $E$'s}.\label{H-cond2}
\eea 
No where had we used  $[H_I,H_J,H_K] =0 $. 
This means our
results obtained in \cite{cat1}
are also valid for a special kind of 
generalized Cartan-Weyl $3$-algebras
whose Cartan 
subalgebra is non-abelian but  satisfies \eq{H-cond1} and
\eq{H-cond2}.  
We observe that when the metric is 
non-degenerate, the conditions \eq{H-cond1} 
and \eq{H-cond2} are not independent. In fact we have:
\begin{lem}
The condition \eq{H-cond2} follows from
\eq{H-cond1} if the metric $g_{IJ}$ is non-degenerate. 
\end{lem}
\begin{proof}
To prove this,
take two arbitrary roots $\a$, $\b$ with $\a+ \b$ being a nonzero root. 
Since $\a
\neq 0$, it means there exists some $H_1, H_2$ such that $\a_{12} \neq 0$,
therefore one can write
$E^\a = [H_1,H_2,E^\a]/\a_{12}$. Denote $\D=[H_I,H_J,H_K]$ for 
arbitrary $I,J,K$. The fundamental identity
$[[\D,H_1, E^\b],H_2,E^\a] =
[[\D,H_2,E^\a],H_1,E^\b] + [\D,[H_1,H_2,E^\a],E^\b]
+[\D,H_1,[E^\b,H_2,E^\a]]$ together with the property \eq{H-cond1} implies
 that 
\be
[\D,E^\a,E^\b]=0, \quad \mbox{for $\a+\b \neq 0$}.
\ee 
Next using 
the invariance of the metric, we have
\be
\la [\D,E^\a,E^{-\a}], H_L \ra = - \la E^{-\a}, [\D,E^\a,H_L] \ra =0, \quad
\mbox{for arbitrary $H_L$}.
\ee 
Moreover, one can show that 
\be
\la [E^\a,E^{-\a},\D], E^\d \ra =0,\quad \mbox{for
arbitrary $\d$}. 
\ee
This is so because we have
$
\la [E^\a,E^{-\a},\D], E^\d \ra $ $= - \la E^\a,[E^\d, E^{-\a},\D] \ra =0
$
for arbitrary $\d \neq \a$;
while for $\d = \a$, we can write 
$
\la [E^\a,E^{-\a},\D], E^\d \ra = - \la E^{-\a},[E^\a, E^\d,\D] \ra =0.
$
Therefore $\la [E^\a,E^{-\a},\D], x \ra =0$ for all $x\in \cA$. Since
the metric is non-degenerate, hence
our claim. 
\end{proof}

As a result,  as long as \eq{H-cond1} holds, 
the explicit results of  classification obtained in section 3 of \cite{cat1}
remain valid. To see more clearly the meaning 
of the condition \eq{H-cond1}, 
let us denote the 
$H$'s which appears on the right hand side of $[H_I,H_J,H_K]$ as
$h^{(0)}$, then \eq{H-cond1} is satisfied if 
\be \label{hhE}
[h^{(0)}, H_J, E^\a]=0, \quad \mbox{for all $E^\a$ and $H_J$}.
\ee
Examining our results obtained in \cite{cat1}, e.g. for the general
index $m$ case, one see that
none of the $u^{(\ri)}$, $H_{\Ih_{(\ri)}}$ or  $H_{\Ih_{(\l)}}$
satisfies the relation \eq{hhE}.
However since $v^{(\ri)}$'s are
central in the algebra, it is obvious that our results in section
3  of \cite{cat1} won't be
altered if $v$'s appears on the right hand side of
$[H_I,H_J,H_K]$. This corresponds to a central extension of the Cartan-Weyl
3-algebra, $[H_I,H_J,H_K] = \sum_\ri L_{IJK}{}^\ri v^{(\ri)}$.

There is also another
straightforward way one can generalize our results  obtained in
\cite{cat1} so that  
it is applicable for 
generalized Cartan-Weyl $3$-algebras
satisfying the
condition \eq{H-cond1}. In our check of the consistency conditions
in appendix A there, 
one could have  started with a larger set of Cartan generators
\be \label{sp-cw1}
\{H_I\} = \{ H_\Ih, H_a, h^{(0)}_\mathfrak{z} \},
\ee
where $h^{(0)}_\mathfrak{z} \in E_0$ 
is an additional set of generators with positive norms and  appear 
on the right
hand side of $[H_I,H_J,H_K]$
\be \label{HHHLH}
[H_I, H_J, H_K] =  \sum_\rl L_{IJK}{}^\rl v^{(\rl)} + 
\sum_\mathfrak{z} K_{IJK}{}^\mathfrak{z} h^{(0)}_\mathfrak{z}
\ee
and satisfies the relation \eq{hhE}. Then one can  immediately see
that, with  the relation 
$[H_I, H_J, H_K] =0$ replaced
with the relation \eq{HHHLH}, 
all our results obtained in the appendix A and in section 3 of 
\cite{cat1} remain valid.
The coefficient $L_{IJK}{}^\rl$ is arbitrary while the coefficient
$K_{IJK}{}^\mathfrak{z}$ is constrained so that the Cartan subalgebra is
nilpotent. We  call this a 
{\it special generalized Cartan-Weyl $3$-algebra}. 

We remark that
in \cite{Jindn}, the general form of metric Lie 3-algebras with a maximally 
isotropic center with arbitrary index is obtained.
In our notation, their results  (equation (6)
there) read
\bea
\;[u^{(\ri)}, u^{(\rj)},  u^{(\rk)}] &=&  
\sum_\rl L_{\ri \rj \rk}{}^\rl v^{(\rl)} + \sum_\d K_{\ri \rj
  \rk}{}^\mathfrak{z}
h_\mathfrak{z}^{(0)}, \label{uu1}\\
\; [u^{(\ri)}, u^{(\rj)}, h_\mathfrak{z}^{(0)}] 
& =& - \sum_\rk K_{\ri \rj \rk \mathfrak{z}} v^{(\rk)}, \label{uu2}\\
\; [H_{\Ih_\ri}, u^{(\rj)}, u^{(\rk)} ] &=& 0\quad \mbox{etc.}
\label{uu3}
\eea
This is in fact a special case of our  
special  generalized Cartan-Weyl $3$-algebras
where the 
relation \eq{HHHLH} is solved with a special form of the
$L_{IJK}{}^\rl$ and $K_{IJK}{}^\mathfrak{z}$ 
giving \eq{uu1}-\eq{uu3}. Therefore this class of Lie 3-algebras
is contained within our framework of generalized Cartan-Weyl $3$-algebras.

Let us now look at the question of having a $\cA_4$ embedding.  
In \cite{cat1} we have shown that the Cartan-Weyl 3-algebras of any 
index $m>0$ generally do not contain a $\cA_4$  subalgebra 
\be \label{A4}
[X^\cP,X^\cQ,X^\cR] = i \e^{\cP \cQ \cR \cS} X^\cS
\ee
and hence these Lie 3-algebras cannot contain
any fuzzy $S^3$ solution. What about Lie 3-algebras 
with maximally  isotropic centers? The answer is again negative. 
To see this, let us write $X^\cP =X^\cP{}_{u^{(\ri)}}
u^{(\ri)} + X^\cP{}_{v^{(\ri)}} v^{(\ri)} + X^\cP{}_\mathfrak{z}
h^{(0)}_\mathfrak{z}+ X^\cP{}_{+\mu} x_+^{(\mu)}   + X^\cP{}_{-\mu}
x_-^{(\mu)} + X^\cP{}_{\a} T^\a$ and note that $u^{(\ri)}$ never appears
on the RHS of the 3-brackets and so $X^\cP{}_{u^{(\ri)}}=0$
immediately.  It follows that the modifications in
\eq{uu1}-\eq{uu3} are not effective.
Therefore there is no nonvanishing solution to \eq{A4} for
the class of metric Lie 3-algebras with a maximally isotropic
centers. This includes the Lorentzian 3-algebra and is consistent with the
fact that these kind of BLG theories describe D2-branes rather than 
uncompactified M2-branes. In the same way one can also check that the 
special  generalized Cartan-Weyl 3-algebras also do not contain any 
$\cA_4$ subalgebra.

It should be clear from the above discussion that the reason why the
Cartan-Weyl 3-algebras or the special generalized Cartan-Weyl 
3-algebras do not admit a $\cA_4$ covariant solution is 
because these Lie 3-algebras all have a set of generators $u^{(\ri)}$ that 
do not appear on the right hand side of the 3-brackets. 
For the generalized Cartan-Weyl 3-algebras, this is no longer the 
case.
Generally, a generalized Cartan-Weyl 3-algebra is 
specified by,  as for a Cartan-Weyl 3-algebra,
a set of 2-form roots $\a$ and the structure constants $g_I(\a,\b)$,
together with the
two new sets of coefficients $L_{IJK}{}^M$ and $c(\a,\b,\g)$.
In some sense, the function $c(\a,\b,\g)$ is the structure that
characterize a genuine Lie 3-brackets. However as we have seen in our
analysis in \cite{cat1}, it cannot be turned on unless the coefficient
$L_{IJK}^H$ is also turned on.

With the activation of $L_{IJK}^H$, it
should be possible to solve \eq{A4} nontrivially and hence allow the 
embedding of $\cA_4$ as subalgebra. If this is really the case, 
the class of generalized Cartan-Weyl 3-algebras will then be the most natural 
extension of the class of semisimple Lie algebras, much the same as 
a semisimple Lie algebra always contain the $su(2)$ as a subalgebra. 
This is of course crucial to the construction of fuzzy sphere solution. 
This issue is under investigation.

\section{Discussions}

In this paper, we have proposed a natural reduction condition which connects 
the Lie algebra which appears in the description of multiple D2-branes and 
the Lie 3-algebra which appears in the description of multiple M2-branes. 
Furthermore, we have shown that this reduction condition implies a great 
simplification of the root space decomposition of the Lie 3-algebras,
leading to the notion of a generalized Cartan-Weyl 3-algebras. A generalized 
Cartan-Weyl 3-algebra is very similar to a Cartan-Weyl 3-algebra, 
but with the important difference that its Cartan subalgebra is 
non-abelian in general.
This is a new feature in the theory of Lie $n$-algebras which does not show up 
for the special case of $n=2$. 

When the Cartan subalgebra is Abelian, a 
complete classification of the Cartan-Weyl 3-algebras has been obtained in
\cite{cat1}. It will be interesting to classify the generalized Cartan-Weyl
3-algebras in general; 
and to work out the specific conditions on $L_{IJK}{}^M$ so that 
one have embedding of the Lie 3-algebra $\cA_4$.

In this paper, we have been mostly concerned with the 
enhanced Lie 3-algebra symmetry
of multiple M2-branes. The description of the enhanced
gauge symmetry of multiple M5-branes is even more mysterious. Taking 
the BLG theory with an infinite dimensional Poisson
bracket as the underlying Lie 3-algebra, 
a new formulation of a single M5 brane has been proposed recently \cite{m2m5}.
The studies of the gauge symmetry in this description and its
resolution may serve to 
teach us something about the enhanced symmetry of
multiple M5-branes, see for example \cite{PT,kazu,peterm5} for some discussions.
Another approach is to study the dynamics of the underlying multiple self-dual
strings of the M5-brane theory using the multiple open M2-branes
system \cite{CS2,Berman}. 
A third approach is through the use of the quantum geometry of
M5-brane in the presence of a constant $C$-field discovered in \cite{CS1}
\footnote{
The supersymmetric coupling of multiple M2-branes to 
a non-constant $C$-field background has been constructed recently in 
\cite{lam}.  
}. 
As noted there, since
the Lie algebraic
structure describing the symmetry of multiple D-branes 
already appeared in the theory of a single
D-brane when  a $B$-field is present, the
quantum geometry of a single  M5-brane in the presence of $C$-field 
should tell us something about the gauge symmetry structure of
multiple M5-branes without $C$-field. Very recently, an understanding
\cite{szabo}
of the quantum geometry of \cite{CS1} has been achieved in terms of a
quantization of the Nambu brackets \cite{nambu}. It will be very interesting
to understand how gauge symmetry is realized in this kind of quantum geometry. 
It will be helpful to be able to derive the Nambu dynamics from a 
fundamental well controlled description \cite{CH}.

Based on the analysis of the entropy counting for M2-branes, it was suggested
in \cite{CHMS} that the Lie 3-algebra which is relevant for the 
description of multiple M2-branes is given by a quantum Nambu bracket. 
Combined with the arguments here, it means the quantum Nambu bracket should
be strong-semisimple. It will be interesting to see if this is  the same as 
the
class of generalized Cartan-Weyl 3-algebras which allows $\cA_4$ embedding.
The quantization of Nambu bracket
is an intriguing problem of great difficulty. See \cite{tak,nambu1,
cubic} for some further discussions on this issue.

\newpage 
\section*{Acknowledgements}

It is a pleasure to thank Douglas Smith for many discussions and 
useful comments.  
The research is supported by EPSRC and STFC.


\end{document}